\documentclass[9pt,twocolumn]{extarticle}
\pdfoutput=1
\usepackage{soul}
\usepackage{titlesec} 
\usepackage[superscript, nomove]{cite}

\usepackage{float}
\usepackage{caption}
\usepackage{lipsum,ulem}
\usepackage[verbose]{placeins}

\usepackage{dsfont}
\usepackage{graphicx}
\usepackage{subcaption}
\usepackage[margin=0.9in]{geometry}
\usepackage[usenames,dvipsnames]{xcolor}
\usepackage[colorlinks,linkcolor=Blue,urlcolor=Blue,citecolor=Blue]{hyperref}
\usepackage{amsmath,amssymb}
\usepackage[squaren]{SIunits}

\usepackage{mathpazo}
\usepackage{courier}
\normalfont
\usepackage[T1]{fontenc}
\usepackage{caption}
\usepackage{upgreek}

\newcommand{\ad}[1]{\textsuperscript{#1}\kern-2pt}

\makeatletter
\def\blx@maxline{77}
\makeatother

\usepackage{capt-of}

\def\mytitle{Toward a quantum resistance standard above liquid nitrogen temperature}

\title{\vspace{-1.0cm}\huge\textbf{\textrm{\mytitle}}}  

\author{Kaining Yang,$^{1}$ Changwei Zhai,$^{2}$ Naijie Ren,$^{3,4}$ Kai Zhao,$^{3,4}$ Chengyuan Li,$^{1}$ Changhao Zhao,$^{2}$ Yunfeng Lu,$^{2}$ \\ Yaning Wang$^{1\dagger}$ Jianting Zhao,$^{2\dagger}$ Zheng Vitto Han,$^{1,3,4\dagger}$}

\date{} 
\begin{document}
\twocolumn[{
\maketitle 
\vspace{-5mm}
\begin{center}
\begin{minipage}{1\textwidth}
\begin{center}
\textit{
\\\textsuperscript{1}Liaoning Academy of Materials, Shenyang 110167, China
\\\textsuperscript{2} Division of Electrical and Magnetic Metrology, National Institute of Metrology, Beijing, People’s Republic of China.
\\\textsuperscript{3} State Key Laboratory of Quantum Optics and Quantum Optics Devices, Institute of Optoelectronics, Shanxi University, Taiyuan 030006, China
\\\textsuperscript{4} Collaborative Innovation Center of Extreme Optics, Shanxi University, Taiyuan 030006, China
\vspace{5mm}
\\{$\dagger$} Corresponding to: ynwang91@gmail.com, zhaojt@nim.ac.cn, vitto.han@gmail.com 
\vspace{5mm}
}
\end{center}
\end{minipage}
\end{center}
\setlength\parindent{13pt}
\begin{quotation}
\noindent 
\section*{Abstract}
{\textbf{Quantum Hall effect (QHE) -- the pioneering demonstration of one-dimensional topologically-protected edge channels -- manifest a transversal resistance quantized solely by two universal constants $e$ and $h$ (the elementary charge and the Planck constant, respectively), referred to as the von Klitzing constant $R_\mathrm{K}=h/e^{2}$. For decades, this has enabled a unique route for realizing quantum Hall resistance (QHR) standard in the \textbf{\textit{Système International d'unités}} (SI), yet high precision QHR under relaxed conditions, especially  temperature higher than 30 K, has not been possible. Here, we show that interfacial charge-order coupling inbetween CrOCl and monolayer graphene offers enhanced Fermi velocity, creating a robust Landau cyclotron gap. Using a customized direct current comparator (DCC, mounted at room temperature), we obtain relative deviation of the Hall resistance from $R_\mathrm{K}/2$ of (15 $\pm$ 23)\,n\textbf{$\Omega$} \textbf{$\Omega$}$^{-1}$ (23 parts per billion of relative uncertainty) at 50 K, with an external magnetic field of 12 T and a dc current of 10 $\mu$A. At 100 K, the system exhibits QHR with 2 parts per million of relative uncertainty.  Our results therefore suggest a platform toward a next-generation quantum resistance standard that may work at above liquid nitrogen temperature and free of conventional compressor refrigerators}.}
\end{quotation}}]
\newpage 
\clearpage

\section*{Introduction}
High mobility two-dimensional electron gas, when subjected to an external perpendicular magnetic field at cryogenic temperatures, often show a quantized Hall resistivity that is expressed by two fundamental constants written as $R_\mathrm{K}=h/e^{2}$ -- known as the famed quantum Hall effect (QHE) \cite{PhysRevLett.45.494,RevModPhys.58.519,Klitzing_1985}. Here $R_\mathrm{K}$ is the so-called von Klitzing constant, and $e$ the elementary charge and $h$ the Planck constant. Ultra-high precision measurement of $R_\mathrm{K}$ has soon been realized with uncertainty reaching 10$^{-10}$, making Quantum Hall systems particularly attractive in regard of resistance standard\cite{BJeckelmann_2001,hartland1992quantum,von2005developments}. Indeed, since the 2019 revision of the base units of the \textit{Système International d'unités}, the unit of electric resistance ($\Omega$) has been defined by $R_\mathrm{K}$\cite{stock2019revision}. 

Nevertheless, several challenges prevail in the field of quantum Hall resistance (QHR). In particular, there is thus far a lack of relaxed experimental condition in terms of eliminating the requirement of ultra high magnetic field and the need of very low temperature. Recently, attempts have been made to tackle the former magnetic field problem by incorporating the quantum anomalous Hall effect (QAHE) into the quantum metrology system at milikelvin temperatures with nanoampere currents \cite{okazaki2022quantum,patel2024zero,gotz2018precision,rodenbach2025unified}. However, the latter ultra-low-temperature limitation remains unresolved so far. 

Since the seminal experimental discovery four decades ago, very few solid state systems including semiconductor quantum wells (GaAs/AlGaAs, etc.), graphene, and two-dimensional semiconductors can manifest QHE\cite{klitzing1980new,PhysRevLett.48.3,novoselov2005two,zhang2005experimental,li2016quantum,zhao2024fractional,delahaye2003revised}. Among them, graphene is of special interest due to its large cyclotron gap between the zero$^{th}$ and first Landau levels\cite{novoselov2007room,giesbers2007quantum}, as well as the programmable quantum Hall plateaux by introducing defects and/or dopants\cite{ju2014photoinduced,wang2022quantum,domaretskiy2025proximity,elias2009control,lu2023synergistic}, or by the coupling of epitaxial substrate that is used to grow graphene in a chemical vapour deposition manner \cite{he2018uniform,janssen2011anomalously,alexander2016giant}. For example, epitaxial graphene on SiC holds great promise in QHR applications\cite{tzalenchuk2010towards,lafont2015quantum,ribeiro2015quantum,couedo2026robust,for2026companion}, since the relative deviation of their Hall resistance from $R_\mathrm{K}$ can readily reach a level of a few parts per billion. The key task still remains in elevating the temperature to realize QHR in graphene -- preferably above a few tens of Kelvin while keeping the precision -- so that a future bench-top compact QHR can be designed free of conventional compressor refrigerators.

In this work, we leverage the interfacial charge-order coupling between CrOCl and monolayer graphene, which creates a greatly enhanced Fermi velocity and thus an enlarged Landau cyclotron gap. Using our customized direct current comparator (DCC, mounted at room temperature), relative deviation of the Hall resistance from $R_\mathrm{K}/2$ of (15 $\pm$ 23) \,n\textbf{$\Omega$} \textbf{$\Omega$}$^{-1}$ (i.e.\ 23 parts per billion, or ppb, of relative uncertainty) at 50 K can be obtained in a typical 0.1 mm-sized hall bar device, with an external magnetic field of 12 T and a dc current of 10 $\mu$A. At $T$=100 K, the system exhibits a larger relative uncertainty of a few parts per million, or ppm. Multiple gate electrodes are used in the measurement paradigm, but negligible leakage current (<10$^{-11}$ A) was found throughout the tests. Our results demonstrate graphene/CrOCl heterostructure as a promising platform for next-generation quantum resistance standards at elevated temperatures, which are potentially distributable, bench-top sized, compressor-free, and quantum-traceable.

\section*{Results and Discussion}
\noindent\textbf{Fabrication and characterizations of 0.1 mm-sized graphene/CrOCl heterostructure Hall bar devices.} 
\\
We fabricated dual-gated Hall bars from h-BN/monolayer graphene/CrOCl van der Waals heterostructures. The constituent flakes were exfoliated from bulk crystals and assembled by dry transfer under ambient conditions. Flake thickness and surface morphology were characterized using a atomic force microscope. We patterned the Hall bars by reactive on etching and fabricated the dual-gated devices using electron-beam lithography. Figure 1a shows a schematic of the graphene/CrOCl heterostructure Hall bar device and its electrical measurement configuration. The cross-sectional schematic in Figure 1c shows the sequence of materials in the device stack. Figure 1b shows the optical image of a typical graphene/CrOCl heterostructure Hall bar device with a lateral span of approximately 0.1 mm.

To describe transport under dual gating, the top and bottom gate voltages are converted to an effective displacement field $\textit{D}_\mathrm{eff}$ and the induced total carrier $n_\mathrm{tot}$, defined as $\textit{D}_\mathrm{eff} = (\textit{C}_\mathrm{tg}\textit{V}_\mathrm{tg} - \textit{C}_\mathrm{bg} \textit{V}_\mathrm{bg}) / 2\epsilon_0 - \textit{D}_\mathrm{0}$, and $\textit{n}_\mathrm{tot} = (\textit{C}_\mathrm{tg} \textit{V}_\mathrm{tg} + \textit{C}_\mathrm{bg} \textit{V}_\mathrm{bg}) / \textit{e} - \textit{n}_\mathrm{0}$. Here, $\textit{C}_\mathrm{tg}$ and $\textit{C}_\mathrm{bg}$ are the top and bottom gate capacitances per unit area, and $\textit{V}_\mathrm{tg}$ and $\textit{V}_\mathrm{bg}$ are the corresponding gate voltages. The offsets $\textit{D}_\mathrm{0}$ and $\textit{n}_\mathrm{0}$ are the residual displacement field and residual doping, respectively. The constants $\epsilon_0$ and $\textit{e}$ denote the vacuum permittivity and elementary charge, respectively.

At  fixed $\textit{D}_\mathrm{eff}$ = 0.69 V/nm, the Hall resistance of Sample-S23 exhibits a plateau that over a broad range of $\textit{n}_\mathrm{tot}$, with the Hall resistance $\textit{R}_\mathrm{H}$ close to the quantized value $\textit{h}/(2e^2)$, as shown in Figure 1d. Within the blue-shaded interval, a robust $\nu$ = -2 Hall plateau is accompanied by a minimum of $\textit{R}_\mathrm{xx}$ $\approx$ 100 m\textbf{$\Omega$}. And the contact resistance of the QHR device is mostly in the tens of Ohms. These conditions define the gate range for subsequent precision measurements of the quantized Hall resistance.

Figure 1e and 1f show maps of the $\textit{R}_\mathrm{xx}$ and Hall conductivity $\textit{$\sigma_\mathrm{xy}$}$, respectively, in the $\textit{n}_\mathrm{tot}$-$\textit{D}_\mathrm{eff}$ space at 1.5 K and a perpendicular magnetic field of 12 T. At negative $\textit{n}_\mathrm{tot}$, the nearly vertical Landau level features indicate that filling factor depends only weakly on $\textit{D}_\mathrm{eff}$, as expected for carriers confined to the graphene plane. On the positive $\textit{n}_\mathrm{tot}$ side, the $\nu$ = -2 and -6 plateaux extend across a broad region accompanied with a suppressed $\textit{R}_\mathrm{xx}$. The CrOCl layer acts as a charge reservoir by exchanging carriers with graphene through its surface states, where electrons may form an electronic crystal, as reported in previous works\cite{wang2022quantum}.

\begin{figure*}[ht!]
	\centering
	\includegraphics[width=0.90\linewidth]{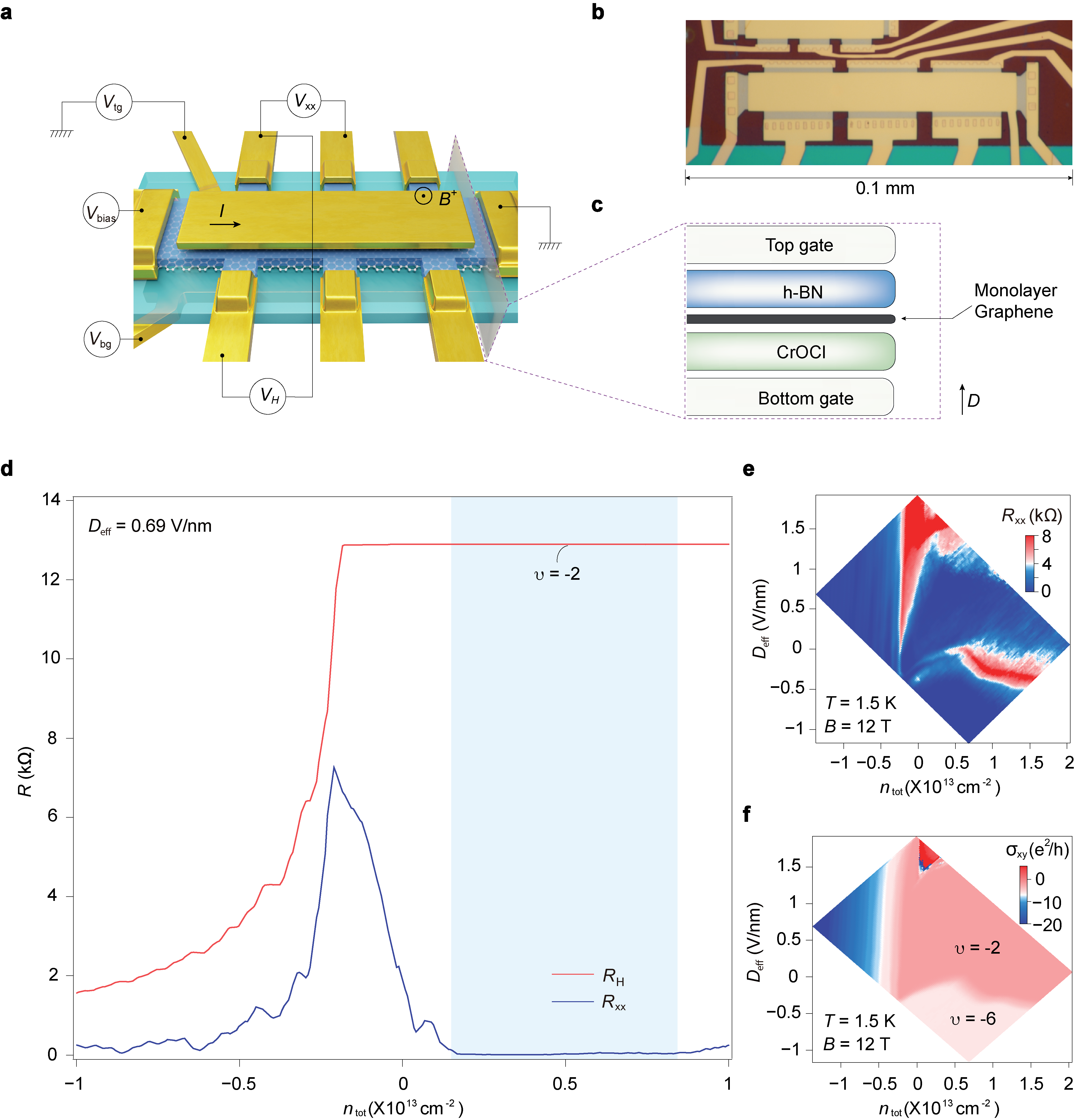}
	\caption{
		\textbf{Device structure and quantum Hall transport in a graphene/CrOCl heterostructure Hall bar device.} \textbf{a}, Schematic of the device and measurement circuit. $\textit{V}_\mathrm{tg}$ and $\textit{V}_\mathrm{bg}$ are the top and bottom gate voltages. The applied bias ($\textit{V}_\mathrm{bias}$) drives a current ($\textit{I}$) between the source and drain. $\textit{V}_\mathrm{xx}$ and $\textit{V}_\mathrm{xy}$ denote the longitudinal and Hall voltages, respectively. The magnetic field ($\textit{B}$) is perpendicular to the device plane.
        \textbf{b}, Optical micrograph image of the representative device. The horizontal marker indicates 0.1 mm.
        \textbf{c}, Schematic depiction of the layered structure of the device. The arrow indicates the positive direction of the displacement field.
        \textbf{d}, Hall resistance $\textit{R}_\mathrm{H}$ (red) and longitudinal resistance $\textit{R}_\mathrm{xx}$ (blue) as functions of $\textit{n}_\mathrm{tot}$ at $\textit{D}_\mathrm{eff}= 0.69\  \textrm{V/nm}$. The blue shading marks the portion of the robust $\textit{$\nu$}= -2\ $ quantum Hall plateau where $\textit{R}_\mathrm{xx}$ remains close to zero.
        Maps of $\textit{R}_\mathrm{xx} (\textbf{e})$ and Hall conductance $\textit{$\sigma_\mathrm{xy}$}$ (\textbf{f})  as functions of $\textit{n}_\mathrm{tot}$ and $\textit{D}_\mathrm{eff}$, measured at $\textit{T}= 1.5\  \textrm{K}$ and $\textit{B}= 12\  \textrm{T}$.
	}
	\label{fig:fig1}
\end{figure*}

 \begin{figure}[ht!]
 	\centering
 	\includegraphics[width=0.95\linewidth]{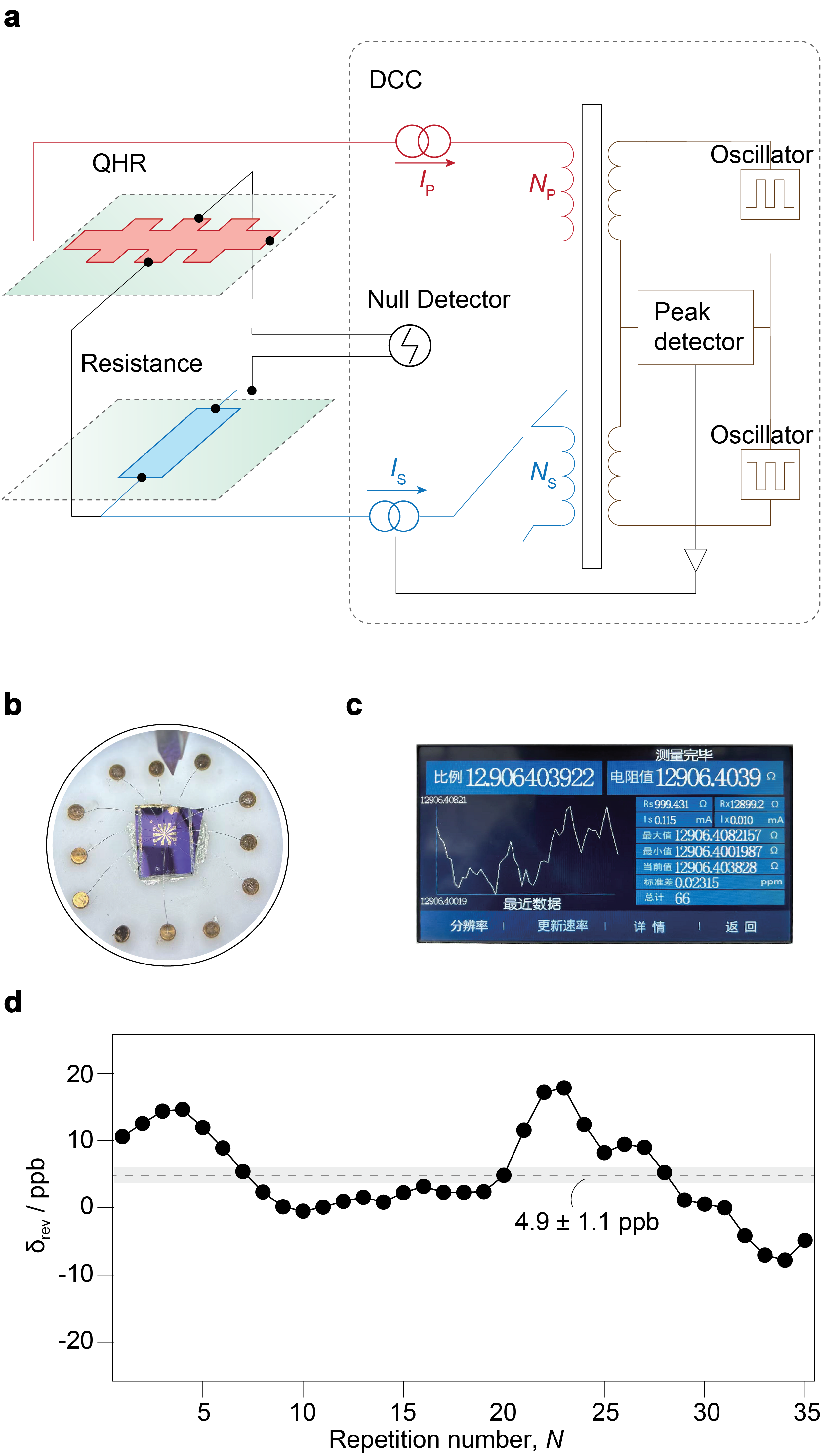}
 \caption{\textbf{DCC set-up and the reversal ratio error.}
 	\textbf{a}, Simplified schematic of the resistance bridge circuit based on DCC. The DCC device was placed at room temperature. The primary current $I_{\mathrm{P}}$ and secondary current $I_{\mathrm{S}}$ were flowed through the QHR device and the 1000 $\Omega$ reference resistor, respectively. The magnetic flux generated by the two coils $N_{\mathrm{P}}$ and $N_{\mathrm{S}}$ was balanced by oscillator and peak detector. The current satisfy the balance condition $N_{\mathrm{P}}$$I_{\mathrm{P}}$ = $N_{\mathrm{S}}$$I_{\mathrm{S}}$ by adjusting feedback control. The voltage difference between QHR and reference resistor was measured by the null detector. When both the magnetic flux and voltage were balanced, the QHR can be precisely measured from the coil numbers and the known value of the reference resistor,  $R_{\mathrm{H}}$$N_{\mathrm{S}}$ = $R_{\mathrm{S}}$$N_{\mathrm{P}}$.
 	\textbf{b}, The optical image of the QHR device bonded on the holder.
 	\textbf{c}, The front panel of DCC measurement, which include resistance ratio, current value, standard deviation and repetition number. 
 	\textbf{d}, The reversal ratio error obtained from a 1 k$\Omega$ : 1 k$\Omega$ resistor comparison at a measurement current of 3 mA. The mean reversal ratio error is 4.9$\pm$1.1 ppb, which demonstrates a ppb level measurement uncertainty of the DCC bridge.
 }
    \label{fig:fig2}
 \end{figure}

\begin{figure*}[ht!]
 	\centering
 	\includegraphics[width=0.95\linewidth]{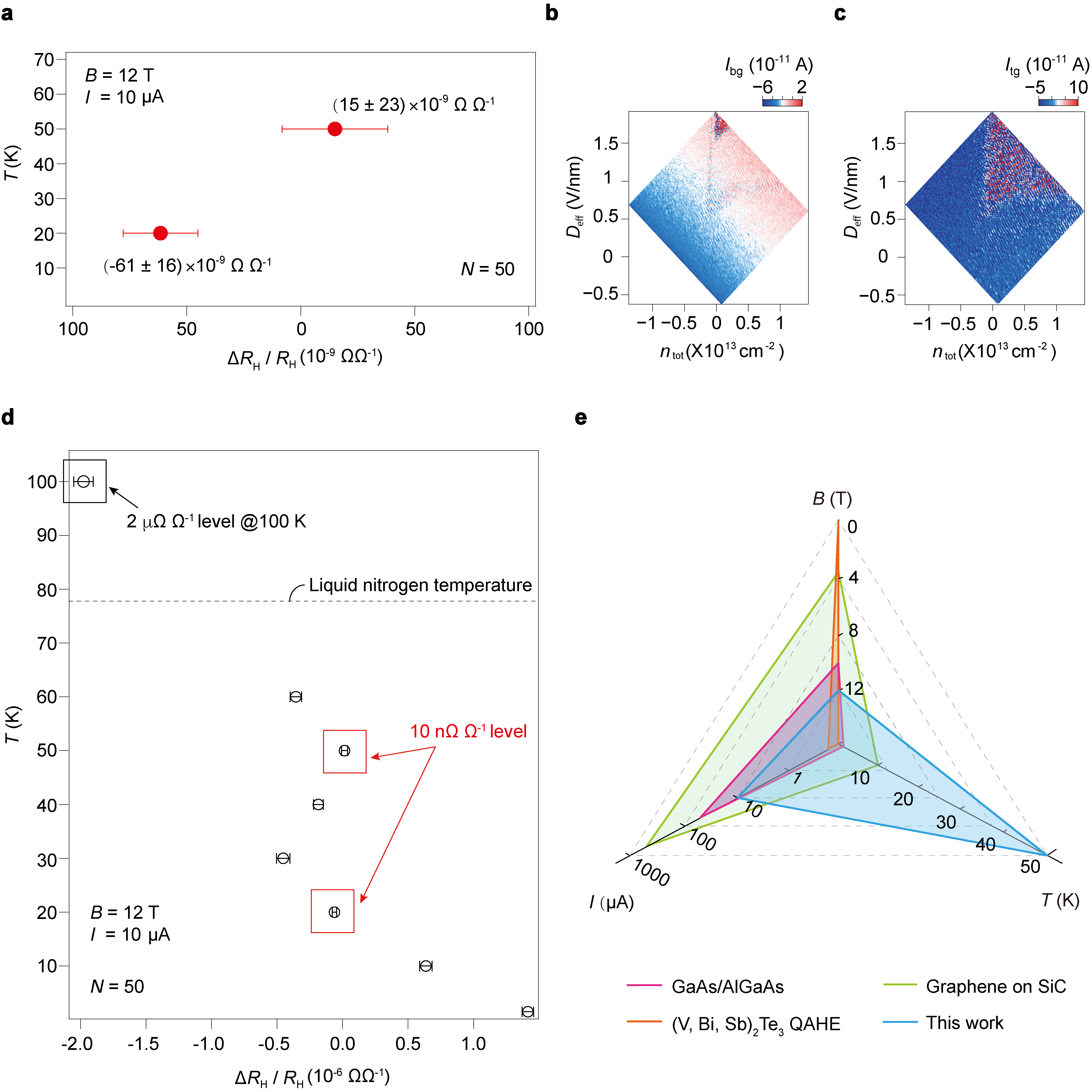}
 	\caption{\textbf{High-precision QHR performances above 50 K in graphene/CrOCl heterostructures.}
	\textbf{a}, The relative deviation $\Delta R_{\mathrm{H}}/R_{\mathrm{H}} = \frac{R_{\mathrm{H}}-R_{\mathrm{K}}/2}{R_{\mathrm{K}}/2}$ of the Hall resistance $R_{\mathrm{H}}$ from the quantized value $R_{\mathrm{K}}$/2 at $T=20$ K and $T=50$ K, obtained utilizing DCC bridge under perpendicular magnetic field $B=12$ T and current $I=10$ $\mu\mathrm{A}$. The error bars indicate the measurement standard deviation, with the number of repetitions $N=50$ at each temperature.
    \textbf{b-c}, Colour maps of the bottom gate leakage current and top gate leakage current as functions of the total carrier density $n_{\mathrm{tot}}$ and the effective displacement field $D_{\mathrm{eff}}$ at $T=50$ K and $B=5$ T. The colour scales are expressed in units of 10$^{-11}$ A.
    \textbf{d}, Temperature dependence of the relative deviation measured at $B=12$ T and $I=10$ $\mu\mathrm{A}$. Red boxes highlight the measurements at $T=20$ K and $T=50$ K, where the deviation reach the 10 n\textbf{$\Omega$} \textbf{$\Omega$}$^{-1}$ level. The deviation increases to approximately 2 $\mu\Omega\,\Omega^{-1}$ at $T=100$ K in the black box. The dashed line marks liquid nitrogen temperature (77 K). Each data point represents $N=50$ measurements.
    \textbf{e}, Comparison of representative magnetic field, current and temperature conditions among GaAs/AlGaAs devices, epitaxial graphene on SiC, V-doped $(\mathrm{Bi},\mathrm{Sb})_2\mathrm{Te}_3$ QAHE devices and the graphene/CrOCl heterostructures at the deviation of around 10 n\textbf{$\Omega$} \textbf{$\Omega$}$^{-1}$ level or below.
    }
 	\label{fig:fig3}
 \end{figure*}

\vspace{5mm}
\noindent\textbf{DCC measurement of QHR in graphene/CrOCl devices.} 
\\

We measured the quantized Hall resistance of the graphene/CrOCl device with a customized room-temperature direct current comparator (DCC) resistance bridge, comparing the device directly with a precision $1000\,\Omega$ standard resistor. Figure 2a shows a simplified schematic of the bridge circuit. The QHR device was connected in series with the primary winding ($N_\mathrm{P}$), and the $1000\,\Omega$ standard resistor in series with the secondary winding ($N_\mathrm{S}$). The bridge was balanced in two steps: the ampere-turn condition $N_\mathrm{P}I_\mathrm{P} = N_\mathrm{S}I_\mathrm{S}$ was first established through the feedback loop, and the residual voltage difference between the QHR device and the standard resistor was then nulled with a nanovoltmeter-based detector. At balance, the Hall resistance is given by $R_\mathrm{H} = R_\mathrm{S}(N_\mathrm{P}/N_\mathrm{S})$, where $N_\mathrm{P}/N_\mathrm{S} \approx 12.9$ for $R_\mathrm{H} \approx R_\mathrm{K}/2$ and $R_\mathrm{S} = 1000\,\Omega$. Current reversal was applied to eliminate thermal EMFs and to cancel the residual gain and offset of the detector. The reversal ratio error of the bridge was $4.9\pm1.1$ ppb in a $1\,\mathrm{k}\Omega:1\,\mathrm{k}\Omega$ comparison at 3\,mA (Figure 2d), confirming the ppb-level performance required here.

The 1000 \,$\Omega$ standard resistor, with a relative uncertainty better than $1\times10^{-8}$, was traceable to the quantum Hall resistance primary standard maintained at National Institute of Metrology (NIM), China. The DCC bridge was calibrated against the NIM cryogenic current comparator (CCC) at the working ratio, and its ratio uncertainty was better than $2\times10^{-8}$. These contributions, together with the measurement repeatability of $2.3\times10^{-8}$ at 50\,K ($N = 50$) and the residual leakage and thermal effects of $1\times10^{-8}$, gave a combined standard uncertainty ($k = 1$) of about $3.4\times10^{-8}$, which was mainly limited by the bridge ratio and the measurement repeatability.

\vspace{5mm}
\vspace{5mm}
\vspace{5mm}

\vspace{5mm}
\noindent\textbf{QHR performances above 50 K in graphene/CrOCl heterostructures.} 
\\

To characterize the robustness of the QHR in graphene/CrOCl heterostructures, we performed high precision Hall resistance measurements over an extended temperature range using a DCC bridge. Figure 3a shows the relative deviations $\Delta R_{\mathrm{H}}/R_{\mathrm{H}}$ of the measured Hall resistance from the theoretically quantized value $R_{\mathrm{K}}$/2 in Sample-S23. At fixed magnetic field $B=12$ T and current $I=10$ $\mu\mathrm{A}$, the relative deviations are $(-61 \pm 16)\times 10^{-9}\,\Omega\,\Omega^{-1}$ at $T=20$ K and $(15 \pm 23)\times 10^{-9}\,\Omega\,\Omega^{-1}$ at $T=50$ K, respectively. Remarkably, at 50 K, the deviation is consistent with the quantized value within the reported uncertainty, with an absolute deviation of only 15 ppb. Figure 3d summarizes the relative deviations from the low temperature up to 100 K under $B=12$ T and $I=10$ $\mu\mathrm{A}$. The small deviations are observed at $T=20$ K and $T=50$ K corresponding to 10$^{-8}$ level, as highlighted in the red boxes.At 100 K, as highlighted in the black box, the QHR device maintains a relative deviation of approximately 2 $\mu\Omega\,\Omega^{-1}$, which demonstrates a robust performance at the temperature exceeding that of liquid nitrogen. It thus manifests great potential for future metrology applications. 

To assess the potential influence of the gate leakage current on measurement accuracy, we characterized the leakage currents through the bottom and top gates as functions of  $n_{\mathrm{tot}}$ and $D_{\mathrm{eff}}$ by employing Keithley 2400 source meter. The corresponding colour maps are presented in Fig. 3b and 3c. Both gates exhibit extremely low leakage currents on the order of 10$^{-11}$ A, suggesting that these weak leakage currents does not introduce an obvious contribution to the high precision resistance measurements.

 Clearly, the current graphene/CrOCl system is the only known system that enters the 10$^{-8}$ level near the liquid temperature. Nevertheless, the temperature dependence does exhibit scattered mean values that deviate from $R_{\mathrm{K}}$/2. Especially, at $T=1.5$ K, the relative deviation increases by an order of magnitude. We suggest that this behavior may arise from uncertainties in the selection of the optimal gate voltages, together with temperature-induced carrier density shift and charge redistribution at the CrOCl interface. All these technical details might be further optimized by improving sample quality, contact resistance, geometric structure, as well as the relative experimental protocols. 

We further compare these results with those of established QHR systems. Figure 3e summarizes the representative operating conditions of the conventional GaAs/AlGaAs 2D electron gases, epitaxial graphene on Silicon carbide (SiC), magnetic topological insulator QAHE devices and the graphene/CrOCl heterostructures in Fig. 3e. These conditions focus on three key experimental parameters: magnetic field, measurement current and temperature. The comparison considers systems exhibiting the relative deviations on the order of 10 n\textbf{$\Omega$} \textbf{$\Omega$}$^{-1}$ level or below. In contrast, the  graphene/CrOCl heterostructures exhibit a distinctive combination of operating temperature and measurement current. Remarkably, at $T=50$ K, $I=10$ $\mu\mathrm{A}$, $B=12$ T, the measured Hall resistance exhibits a relative deviation of $(15 \pm 23)\times 10^{-9}\,\Omega\,\Omega^{-1}$, highlighting the potential of interfacial charge coupling for extending the operating temperature range of precision quantum Hall resistance measurements.

It is known that pristine graphene possesses a distinct band structure and the resulting N$^{th}$ Landau level (LL) at the energy of $\varepsilon_\mathrm{LL}(N)=v_{F}\mathrm{sgn}(N)\sqrt{2e\hbar B|N|}$ under a fixed magnetic field $B$. Nevertheless, in the scenario of the graphene/CrOCl heterostructure, the key effect of CrOCl on graphene is that the interfacial charge order reconstructs the band structure of ``ordinary'' monolayer graphene \cite{wang2022quantum,lu2023synergistic}, yielding a sharper Dirac cone and a significantly reduced cyclotron mass $m^{*}$ near an order of magnitude smaller than its original value. It thus gives rise to a few times higher Fermi velocity of graphene in the so-called charge-transfer-coupling regime, supported by both Shubnikov–de Haas oscillation measurements and the numerical simulations in our previous studies \cite{wang2022quantum,lu2023synergistic}. Meanwhile, it is believed that, in a metrologically precise level of QHR, the system often requires a Landau Cyclotron gap about 100 times larger than the thermal energy $k_{B}$T \cite{CRPHYS_2019__20_1-2_92_0}. Thus, in our graphene/CrOCl system, the first LL gap $\Delta=v_{F}\sqrt{2\hbar eB}$ will be in the order of about 600 $\sim$ 700 meV at 12 T, which qualitatively depicts a possible quantum metrology application above the liquid nitrogen temperature (77 K, corresponding to $\sim$ 6.4 meV in terms of thermal energy), although there is still improvements to be made in the future to further lower the magnetic field needed.

\vspace{5mm}
To conclude, we have devised a CrOCl/monolayer-graphene heterostructure equipped with dual gate architecture and a lateral size of about 0.1 mm, which creates a greatly enhanced Fermi velocity and thus an enlarged Landau cyclotron gap in the system. A direct current comparator mounted at room temperature was adopted to evaluate the quantum Hall resistance at different temperatures and magnetic fields. Systematic measurements reveal a relative deviation of the Hall resistance from $R_\mathrm{K}$/2 of (15 $\pm$ 23) \,n\textbf{$\Omega$} \textbf{$\Omega$}$^{-1}$ at 50 K can be obtained, with an external magnetic field of 12 T and a dc current of 10 $\mu$A applied to the Hall bar device. At even elevated temperature of $T$ = 100 K, the graphene/CrOCl heterostructure exhibits relative uncertainty at the level of $10^{-6}$.
During the tests, the gate leakage current was found to be on the order of 10$^{-11}$ A or below, ruling out parasitic gate-induced noise in such a dual-gated device structure. The 10 $\mu$A current load, operating temperature above 50 K, together with the few tens of ppb uncertainty in the quantum Hall resistance relative to $R_\mathrm{K}/2$, hallmark the graphene/CrOCl system as an outstanding platform for next-generation bench-top-sized compressor-free, and quantum-traceable resistance standards.

 \clearpage

\clearpage

\section*{\label{sec:level1}Data Availability}

The data that support the findings of this study will be available with a Zenodo doi link when published.

\section*{\label{sec:level2}Code Availability}

The code that supports the findings of this study is available upon reasonable request to the corresponding authors.

\section*{\label{sec:level3}Acknowledgements}
This work is supported by the National Natural Science Foundation of China (NSFC) (Grant Nos. 92265203, 92565302, 12104462, 11974357, and 12304237).

\section*{Author Contributions}
Z.V.H., J.Z. and Y.W. conceived the experiment and supervised the overall project. K.Y. and Y.W. performed the device fabrications. K.Y., C.Z. (Changwei Zhai), N.R., K.Z., C.L., Y.W., and Y.L. performed the electrical measurements. The manuscript was written by Z.V.H., Y.W., K.Y. C.Z.(Changhao Zhao), and J.Z. with discussions and inputs from all authors.

\section*{Competing Interests}
The authors declare no competing interests.

\end{document}